\documentclass[aps,pra,twocolumn,superscriptaddress]{revtex4-2}
\usepackage[utf8]{inputenc}
\usepackage{soul}
\usepackage{textcomp}
\usepackage[english]{babel}
\usepackage{graphicx}
\usepackage{dcolumn}
\usepackage{bm}
\usepackage{amsmath}
\usepackage{verbatim}     
\usepackage[dvipsnames]{xcolor}       
\usepackage[normalem]{ulem}   
\usepackage{bbold}
\usepackage{mathrsfs}
\usepackage[mathscr]{eucal}
\usepackage{physics}
\usepackage{algorithmic}
\usepackage{hyperref}
\hypersetup{
   pdfpagemode=None,
   pdfstartpage=1,
   pdfmenubar=true,
   pdftoolbar=true,
   colorlinks = true,
   linkcolor=blue,
   citecolor=blue,
   urlcolor=blue,
   bookmarksopen=false
 }

\begin{document}
\title{Fourier Quantum Process Tomography}

\author{Francesco Di Colandrea}
\email[Correspondence email address: ]{francesco.dicolandrea@uottawa.ca}
\affiliation{Nexus for Quantum Technologies, University of Ottawa, K1N 5N6, Ottawa, ON, Canada}

\author{Nazanin Dehghan}
\affiliation{Nexus for Quantum Technologies, University of Ottawa, K1N 5N6, Ottawa, ON, Canada}

\author{Alessio D'Errico}
\affiliation{Nexus for Quantum Technologies, University of Ottawa, K1N 5N6, Ottawa, ON, Canada}
    
\author{Ebrahim Karimi}
\affiliation{Nexus for Quantum Technologies, University of Ottawa, K1N 5N6, Ottawa, ON, Canada}
\begin{abstract} 
The characterization of a quantum device is a crucial step in the development of quantum experiments. This is accomplished via Quantum Process Tomography, which combines the outcomes of different projective measurements to deliver a possible reconstruction of the underlying process. The tomography is typically performed by processing an overcomplete set of measurements and extracting the process matrix from maximum-likelihood estimation. Here, we introduce a new technique, referred to as Fourier Quantum Process Tomography, which requires a reduced number of measurements, and benchmark its performance against the standard maximum-likelihood approach. Fourier Quantum Process Tomography is based on measuring probability distributions in two conjugate spaces for different state preparations and projections. Exploiting the concept of phase retrieval, our scheme achieves a complete and robust characterization of the setup by processing a near-minimal set of measurements. We experimentally test the technique on different space-dependent polarization transformations, reporting average fidelities higher than 90\% and significant computational advantage.
\end{abstract}
\maketitle
%
%
%
\date{\today} 

\section{Introduction}
The functionalities of a \emph{black-box} quantum device can be assessed via Quantum Process Tomography (QPT) techniques. These techniques prescribe a set of experimental measurements to identify the unknown parameters of the underlying process matrix~\cite{Chuang1997}. QPT is routinely performed across various quantum architectures, ranging from nuclear magnetic resonances~\cite{PhysRevA.64.012314} to cold atoms~\cite{PhysRevA.72.013615}, trapped ions~\cite{PhysRevLett.92.220402,PhysRevLett.97.220407}, superconducting circuits~\cite{PhysRevB.82.184515,Bialczak2010} and photonic setups~\cite{PhysRevLett.91.120402,Altepeter2003,PhysRevLett.93.080502,Lobino2008, Bongioanni2010,Rahimi-Keshari2013,Ndagano2017,Anton2017,PhysRevA.98.052327,Bouchard2019,DiColandreaQPT}. 

In principle, one could extract the analytical relations between the operator parameters and the outcomes of suitable projective measurements~\cite{LeRoy-Brehonnet1997}. However, this proves to be often incompatible with realistic experimental noise, typically yielding nonphysical reconstructions. This inconvenience can be overcome by formulating the process tomography as an optimization problem, as first proposed for the tomography of quantum states~\cite{James2001}. 

In this framework, the most elementary scenario is the characterization of an SU(2) gate $\hat{U}$ acting on a two-level quantum system (qubit). Polarization of photons provides a natural way of encoding qubits, with $\hat{U}$ implemented via one or multiple birefringent waveplates. Accordingly, the characterization of devices acting on light polarization can be accomplished via QPT~\cite{Aiello2006}.

Here, we address the more challenging scenario of characterizing optical SU(2) gates that are dependent on some $d$-dimensional degree of freedom, hereafter referred to as \textit{lattice}. We introduce a new technique, named Fourier Quantum Process Tomography (FQPT), that allows retrieving all the parameters of the unknown transformation by processing only three sets of projective measurements collected in 2 conjugate planes. This method applies to SU$(2\times d)$ transformations, which can be decomposed in a $2\times 2$ block-diagonal form. 

FQPT is validated experimentally on complex polarization transformations realized via liquid-crystal metasurfaces (LCMSs) patterned with high spatial frequencies~\cite{DiColandrea2023}, and its performance is compared with a standard maximum-likelihood (ML) approach. In this experiment, the measurements can be conveniently chosen to be performed in two conjugate planes, namely the near and far field, wherein the light distributions are directly connected via a Fourier transform. If the near field is associated with an intermediate plane or two intermediate planes are selected, then the Fourier transform is replaced by a paraxial Fresnel propagator~\cite{goodman2005introduction}. FQPT can also be implemented in other platforms. For instance, integrated photonic technologies~\cite{wang2020integrated} can support additional chips specifically implementing the Quantum Fourier Transform (QFT) algorithm~\cite{politi2009shor, crespi2016suppression,flamini2018observation}. At the same time, SU(2) operations could be implemented either in the polarization~\cite{sansoni2012two,heilmann2014arbitrary,pitsios2017geometrically} or path encoding. In the latter case, the waveguide array would simulate a composite lattice. Similar schemes have also been reported in several non-photonic platforms~\cite{weinstein2001implementation,fowler2004implementation,mariantoni2011implementing}.

\section{Theory}
A qubit rotation of an angle 2$E$ around the axis ${\bm{n}=(n_1,n_2,n_3)}$, with $0\leq E<\pi$ and $\abs{\bm{n}}=1$, is described by an SU(2) operator
\begin{equation}
\hat{U}=e^{-iE\bm{n}\cdot\boldsymbol{\sigma}}=\cos(E)\sigma_0-i\sin(E)(\bm{n}\cdot\bm{\sigma}),
\end{equation}
where $\sigma_0$ is the 2$\times$2 identity matrix and $\bm{\sigma}=(\sigma_1,\sigma_2, \sigma_3)$ is the vector of the three Pauli matrices. 


The characterization of an optical SU(2) gate is typically performed by processing an overcomplete set of 16 projective measurements of the form
\begin{equation}
    I_{ab}=\abs{\mel{b}{\hat{U}}{a}}^2,
\end{equation}
where $\ket{a}$ and $\ket{b}$ are extracted from the three sets of states forming the Mutually Unbiased Bases (MUB) of SU(2)~\cite{Chuang1997,PhysRevLett.91.120402,PhysRevLett.93.080502}. The process tomography of the gate is then accomplished via an ML approach, i.e., by minimizing a cost function expressing the distance between the experimental outcomes $I_{ab}^{\text{exp}}$ and the corresponding theoretical predictions $I_{ab}^{\text{th}}$ \cite{PhysRevLett.93.080502,Aiello2006}:
\begin{equation}
    \mathcal{L}=\sum_{ij} (I_{ab}^{\text{exp}}-I_{ab}^{\text{th}})^2.   
\end{equation}
This approach may become unfavourably time-consuming and less accurate in the case of transformations acting on high-dimensional Hilbert spaces. Here, we consider the case of unitaries, which depend on some additional parameter, e.g., a spatial variable or a lattice position. More precisely, we assume the parameters $E,\, n_1,\,n_2,\, n_3$ to be functions of $\mathbf{r}$, where $\mathbf{r}$ can be any set of either discrete or continuous variables. In this current study, we assume position $\mathbf{r}$ as a continuous variable. However, the discrete case can be obtained by replacing integrals with summations, i.e., $\int f(x)\,dx\rightarrow\sum_x\,f(x)$. 

The unknown unitary process acts on quantum states whose Hilbert space is the tensor product of a qubit space $\mathcal{H}_i$, associated with an internal degree of freedom, and a high-dimensional space $\mathcal{H}_r$, associated with the lattice. 

We consider an input state uniformly distributed along $\mathbf{r}$, with an internal state prepared as one of the MUB states $\ket{a}$. $\ket{a}$ is assumed to be the positive eigenvector of $\sigma_1, \sigma_2$ or $\sigma_3$. The unknown unitary $U(\mathbf{r})$ acts on this state, and then a projection onto the same internal state is performed. The probability distribution along $\mathbf{r}$ is ${I_{aa}(\mathbf{r})=\abs{\mel{a}{U(\mathbf{r})}{a}}}^2$. Another measurement is performed to retrieve the probability distribution $\tilde{I}_{aa}(\mathbf{k})$ in the reciprocal space of $\mathbf{r}$. This is enabled by a QFT of the final state:
\begin{equation}
    \tilde{I}_{aa}(\mathbf{k}) = \abs{\int \text{d}^2r\,\bra{a}U(\mathbf{r})\ket{a}e^{i\mathbf{k}\cdot\mathbf{r}}}^2.
\end{equation}
%

\begin{figure*}
\centering
\includegraphics[width=0.8\textwidth]{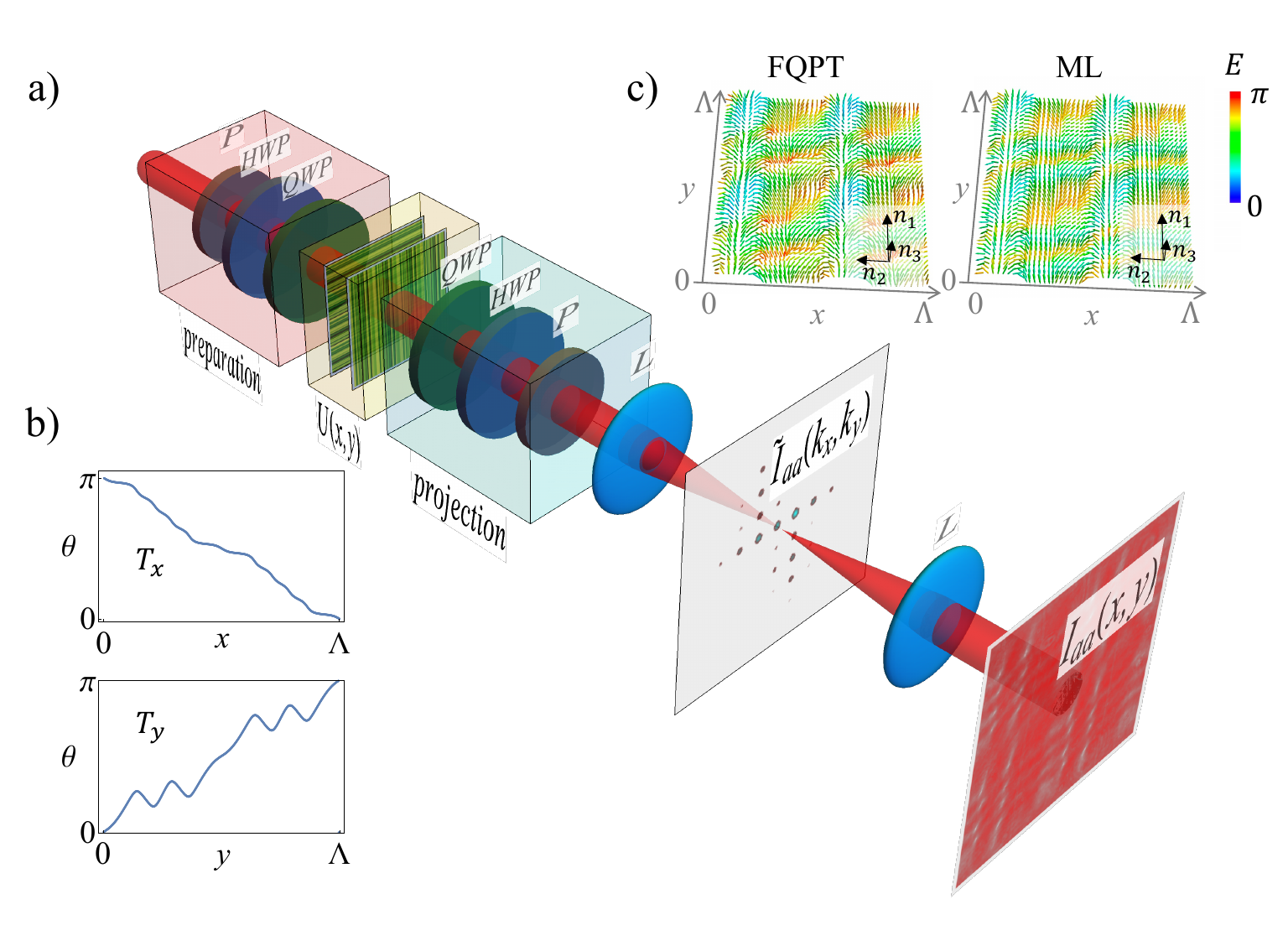}
\caption{\textbf{Fourier Quantum Process Tomography.} a) A space-dependent polarization transformation $U(x,y)$ is implemented via LCMSs. The process tomography is performed by preparing and projecting onto MUB states. 
The resulting intensity distributions are measured in the far field, $\tilde{I}_{aa}(k_x,k_y)$, and image plane, $I_{aa}(x,y)$, of the optical operator. b) Plots of the optic-axis pattern of two plates used to test the technique. These are patterned along orthogonal directions to create a complex 2D modulation. c) Reconstruction of the process $U(x,y)=T_x(x)T_y(y)$ using FQPT and the traditional ML technique. The arrows represent the local eigenvectors $\bm{n}(x,y)$ and their color is associated with the local rotation angle $E(x,y)$ (see also Fig.~\ref{fig:HH-2D}d) for details).}
    \label{fig:setup}
\end{figure*}

One can extract the wave function phase in each plane from the two probability distributions by applying phase-retrieval techniques. We specifically employed the Gerchberg–Saxton (GS) algorithm~\cite{gerhberg1972practical}. In particular, we can retrieve the amplitude ${A_a(\mathbf{r})=\sqrt{I_{aa}(\mathbf{r})}}$ and the phase $\alpha_a(\mathbf{r})=\arg(\mel{a}{U(\mathbf{r})}{a})$. However, the latter is determined up to an unknown constant $\xi_a$. The amplitude and phase are related to the process parameters according to
\begin{eqnarray}
    \mel{a}{U(\mathbf{r})}{a}&=&A_a(\mathbf{r})e^{i\alpha_a(\mathbf{r})+i\xi_a}\\\nonumber
    &=&\cos{E(\mathbf{r})}-i\, n_{a}(\mathbf{r})\sin{E(\mathbf{r})},    
\end{eqnarray}
where ${a=1,2,3}$, depending on which Pauli matrix is considered ($\sigma_a\ket{a}=\ket{a}$). Thus, we obtain
\begin{subequations}
\begin{align}
E(\mathbf{r}) &= \arccos{\left(A_a(\mathbf{r})\cos{(\alpha_a(\mathbf{r})+\xi_a)}\right)}; \label{eq:Ehh} \\
n_a(\mathbf{r}) &= -A_a(\mathbf{r})\dfrac{\sin{(\alpha_a(\mathbf{r})}+\xi_a)}{\sin{E(\mathbf{r})}}. \label{eq:nx}
\end{align}
\end{subequations}


Equations \eqref{eq:Ehh}-\eqref{eq:nx} show that the extracted parameters depend on the global phase shifts $\xi_a$, which cannot be estimated from the phase retrieval method. Indeed, any phase that differs from $\alpha_a(\mathbf{r})$ by a constant global shift yields the same measured amplitude in the direct and reciprocal space. 
Considering the ambiguity due to the global phase shift, we list all the possible energy modulations compatible with the measurements. In practice, we select $N$ values of the global phase shift $\xi_{a,j}=2\pi j/N$,
with ${j=0,1,...,N-1}$. We specifically set ${N=64}$.
Only one of the candidates can best describe the process under investigation. To find it, we perform an additional measurement in the reciprocal space, obtained by evolving any input state without projection on the internal degree of freedom, e.g., 
\begin{equation}
    \tilde{I}_0(\mathbf{k})=\abs{\int \text{d}^2r\,U(\mathbf{r})\ket{b}e^{i\mathbf{k}\cdot\mathbf{r}}}^2,
\label{eq:farfield}
\end{equation}
where $\ket{b}$ can be chosen arbitrarily.
Crucially, this last measurement also provides the normalization factor for all the data. In this way, by numerically simulating the far field obtained from each of the $N$ possible SU(2) evolutions, we can isolate the realization associated with the physical setup by sifting the one that minimizes the distance with the measurement of Eq.~\eqref{eq:farfield}. 

\begin{figure*}[t!]
\centering
\includegraphics[width=1\textwidth]{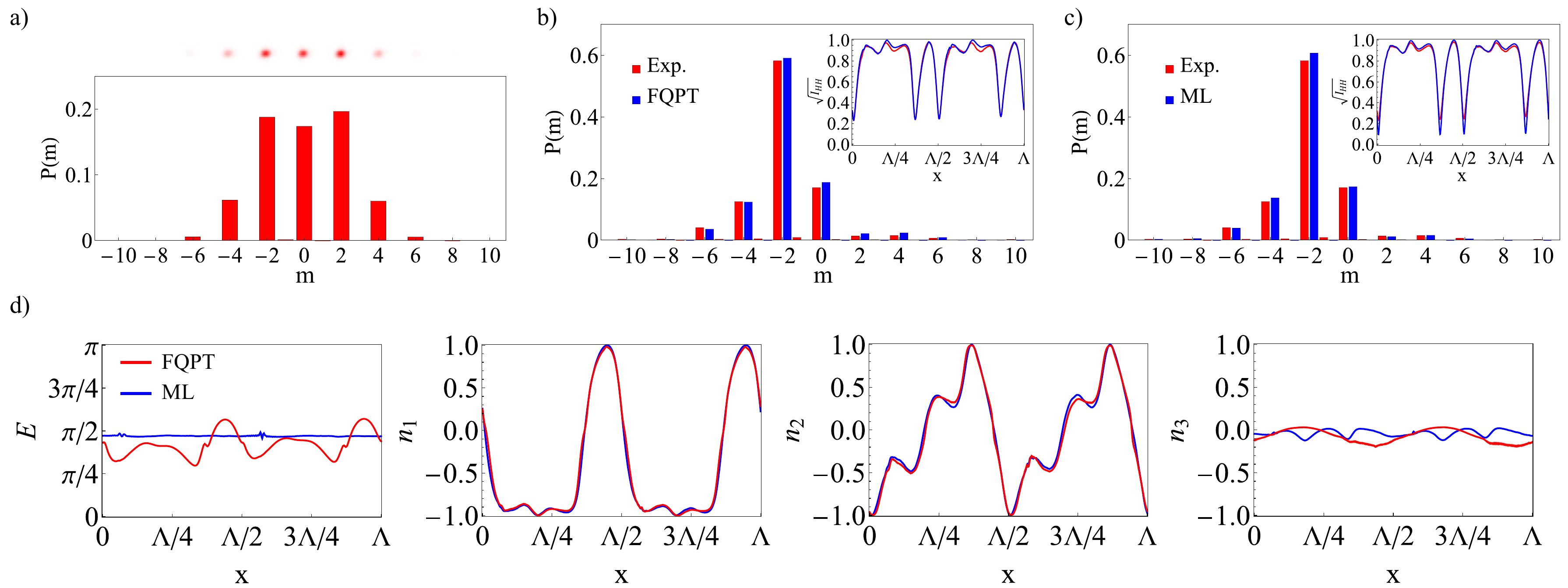}
\caption{\textbf{One-dimensional complex polarization transformations.} a) Far-field intensity distribution $\tilde{I}_{HH}(k_x)$ for a 1D periodic SU(2) transformation. A power spectrum $P(m)$ is extracted after discretizing the reciprocal space. b)-c) Comparison between the experimental total far-field distribution for a $\ket{L}$ input state and the reconstruction obtained from FQPT b) and ML c). The insets show the reconstructed near-field amplitude $\sqrt{I_{HH}(x)}$, compared with the experimental measurement. d) Reconstructions of the process parameters across one period performed by FQPT (red curves) and ML (blue curves). The expected periodicity is captured by both approaches.}
    \label{fig:HH-1D}
\end{figure*}

\begin{figure*}
\centering
\includegraphics[width=0.8\textwidth]{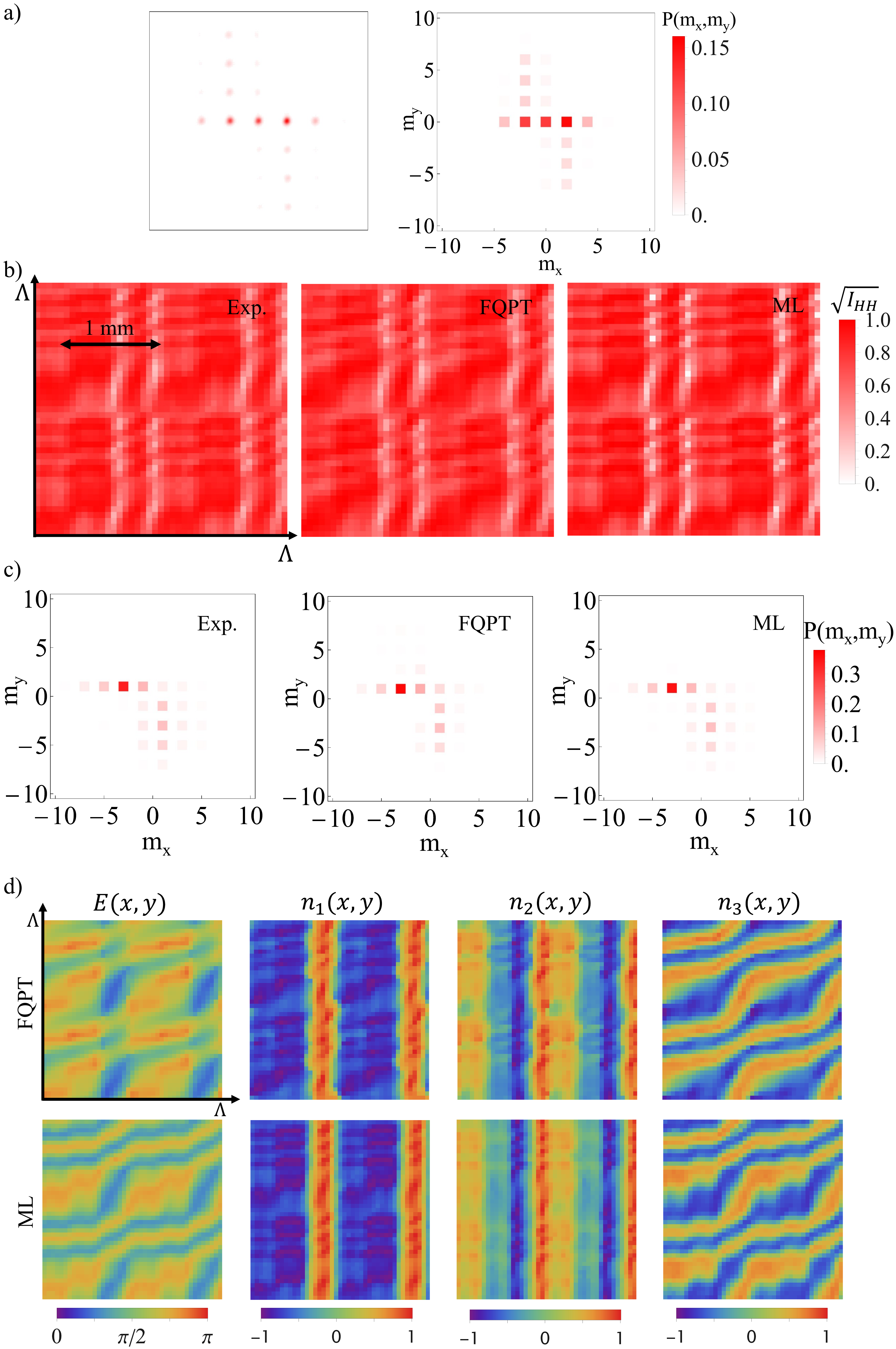}
\caption{\textbf{Two-dimensional complex polarization transformations.} a) Far-field intensity distribution $\tilde{I}_{HH}(k_x,k_y)$ for a 2D periodic SU(2) transformation. A power spectrum $P(m_x,m_y)$ is extracted. b) Comparison between the total far-field distribution for a $\ket{L}$ input state and the reconstruction obtained from FQPT and ML. c) Reconstructed near-field amplitude $\sqrt{I_{HH}(x,y)}$, compared with the experimental measurement. d) Reconstructions of the process parameters across one period performed by FQPT and ML. The expected periodicity is captured by both approaches.}
    \label{fig:HH-2D}
\end{figure*}

We finally remark that, in principle, the method also works with only 2 sets of measurements. For instance, one could process $I_{11}(\mathbf{r})$ and $I_{22}(\mathbf{r})$, together with $\tilde{I}_{11}(\mathbf{k})$ and $\tilde{I}_{22}(\mathbf{k})$, to retrieve $n_1(\mathbf{r})$ and $n_2(\mathbf{r})$. In this case, the third component could be directly computed from the normalization condition: $n_3=\pm\sqrt{1-n_1^2-n_2^2}$. Combined with the final measurement in the reciprocal space, these add up to a minimal set of 5 total measurements, in perfect agreement with the argument provided in Ref.~\cite{DiColandreaQPT}. However, we observed that experimental noise may yield nonphysical results, i.e., $n_3$ features a non-zero imaginary part at some location. This suggested integrating the minimal setting with an additional set of measurements~\cite{James2001,Aiello2006,DiColandreaQPT}.

When accounting for the noise, the proposed technique thus requires only 7 measurements instead of the conventional 16: 3 measurements in the lattice space, 3 in the reciprocal space, and a last one, still in the reciprocal space, to fix normalization and remove all the ambiguities on the parameters of the unitary.  

\section{Experimental Results}
In photonic setups, a qubit can be encoded into photon polarization, which is typically manipulated via optical waveplates. In the circular polarization basis, where $\ket{L}=(1,0)^T$ and $\ket{R}=(0,1)^T$ are left and right circular polarization states, respectively, a waveplate $L_{\delta,\theta}$ having the birefringence $\delta$ and the optic axis oriented at $\theta$ with respect to the horizontal direction can be expressed in the matrix form \begin{equation}L_{\delta,\theta}=\begin{pmatrix}\cos{(\delta/2)} & i\sin{(\delta/2)}e^{-2i \theta}\\i\sin{(\delta/2)}e^{2i \theta} & \cos{(\delta/2)}\end{pmatrix}.\label{eqn:su2matrix}\end{equation}
A single waveplate thus implements a rotation of an angle $-\delta/2$ around the equatorial axis ${\bm{n}=(\cos{2\theta},\sin{2\theta},0)}$. Nevertheless, one can cascade multiple waveplates to implement more general operations \cite{SIMON1990165,Sit_2017}.

We apply FQPT to periodic polarization transformations induced by complex LCMSs~\cite{DiColandrea2023}. These can be modeled as optical waveplates having patterned optic-axis modulation $\theta=\theta(x,y)$ and fixed, but tunable, birefringence~\cite{Rubano2019}. In particular, we test our method with LCMSs featuring high spatial-frequency modulations along the $x$ and $y$ directions. This scenario is experimentally more challenging than the one addressed in Ref.~\cite{DiColandreaQPT}, where simple combinations of polarization gratings were considered. We benchmark our technique against the standard ML approach processing a whole set of 16 polarimetric measurements (all taken in the near field), taking into account both the timing and the accuracy of the final reconstruction. For the minimization, we employed the \emph{NMinimize} routine from Wolfram Mathematica~\cite{nminimize}. 

The experiments are realized with the setup sketched in Fig.~\ref{fig:setup}a). A Ti:Sa laser (wavelength ${\lambda=810}$ nm) is coupled to a single-mode fiber. The output Gaussian mode is magnified with a telescope lens system, $f_1=125 \text{ mm}$ and $f_2=200$ mm (not shown in the figure). The beam waist is measured to be $2.6\pm 0.1 \text{ mm}$. In this way, the overall beam size is larger than the largest periodicity on the plates, that is $\Lambda=2.5\text{ mm}$. A combination of a linear polarizer ($\text{P}_1$), a half-wave plate ($\text{HWP}_1$) and a quarter-wave plate ($\text{QWP}_1$) is needed to prepare any input polarization state. The beam then propagates through one or multiple LCMSs, engineering a space-dependent SU(2) optical operator. Another set $\text{QWP}_2$-$\text{HWP}_2$-$\text{P}_2$ is adjusted to project onto any state. The last element ($\text{P}_2$) is removed when performing the measurement of Eq.~\eqref{eq:farfield}. Each polarimetric measurement is collected on a CCD camera, placed either after a 4$f$ system (${f=150\text{ mm}}$) or in the focal plane of a lens (${f'=250\text{ mm}}$), depending on if the measurement is realized in the near field or in the far field, respectively. Recall that the far-field light distribution is proportional to the transverse momentum distribution, i.e., to the Fourier transform of the input field. 

The first experiment is realized with a single LCMS displaying a complex periodic modulation along the $x$ axis (see Fig.~\ref{fig:setup}b)). The optical retardation is set at $\delta=\pi$. The periodicity of the process simplifies the analysis, as only a discrete spectrum of Fourier components is expected to appear in the far field for all the measurements. 
Furthermore, in this one-dimensional (1D) realization, the intensity modulations can also be integrated along the $y$ direction to mitigate experimental imperfections. 

Figure~\ref{fig:HH-1D}a) shows the far-field intensity distributions recorded for the $\mel{H}{U}{H}$ configuration, with ${\ket{H}=(\ket{L}+\ket{R})/\sqrt{2}}$. As discussed above, the far-field distribution can be discretized and a normalized power spectrum $P(m)$ is extracted. Figures~\ref{fig:HH-1D}b)-c) illustrate the comparison between the experimentally measured total far-field distribution for a $\ket{L}$ input state (see Eq.~\eqref{eq:farfield}) and the reconstruction performed via FQPT and ML, respectively. The agreement with the experimental observation is quantified in terms of the similarity estimator ${s=(\sum_m \sqrt{P_{\text{exp}}(m)P_{\text{rec}}(m)})^2}$, where $P_{\text{exp}}(m)$ and $P_{\text{rec}}(m)$ are the (normalized) experimental and reconstructed far-field distributions. We obtain ${s_{\text{FQPT}}=97.2\%}$ and ${s_{\text{ML}}=97.0\%}$. Another metric that can be considered is the absolute distance between the two distributions, computed as ${\Delta=(\sum_m\abs{P_{\text{exp}}(m)-P_{\text{rec}}(m)})^2}$. We report ${\Delta_{\text{FQPT}}=0.088}$ and ${\Delta_{\text{ML}}=0.083}$. In both figures, the insets show the same comparison for the 1D near-field amplitude ${\sqrt{I_{HH}(x)}}$. Reconstructions of individual parameters are plotted in Fig.~\ref{fig:HH-1D}d). The agreement between the predictions of the two methods is quantified in terms of the fidelity ${F=\abs{\Tr(U_\text{ML}^{\dag}U_\text{FQPT})}/n}$, where ${n=2}$ is the dimension of the internal degree of freedom~\cite{Wang2009}. An excellent average fidelity is obtained, $\bar{F}=95.7\%$, where $\bar{F}$ denotes the average fidelity computed over all the pixels.  
These results prove that both methods provide reliable reconstructions. It must be noted that FQPT achieves satisfactory reconstructions by only processing a near-optimal set of measurements. Moreover, a brute-force minimization approach tends to jump between the parameters associated with processes $U$ and ${e^{i\pi}U=-U}$, as these both generate the same experimental outcomes~\cite{Flaschner2016,Tarnowski2019,DiColandreaQPT,PhysRevResearch.5.L032016}. For this reason, a continuity constraint must be enforced between consecutive pixels. Conversely, our method is assumption-free as it does not rely on any a-priori hypothesis on the process parameters. The technique is also extremely efficient on the computational level. The total times required for a complete reconstruction are ${t_{\text{FQPT}}\approx 1\text{ min}}$ and ${t_{\text{ML}}\approx 30\text{ min}}$. 

Recall that the GS algorithm is executed to retrieve the unknown phases. This algorithm is based on an iterative strategy that presents intrinsic limitations~\cite{Fienup:82}. The convergence speed is sensitive to the initial guess of the phase, and the convergence to a global minimum is not guaranteed. Moreover, the presence of noise in the input data can severely affect the accuracy of phase retrieval. To overcome these limitations, for each set of polarimetric measurements, the algorithm is run ${N_T^{\text{1D}}=100}$ times for ${N_I^{\text{1D}}=1000}$ iterations, randomly initializing the phase guess at each trial. In doing so, the best phase reconstruction can be selected (up to global shifts) as the one minimizing the total distance between the reconstructed and measured near-field amplitudes. 



The second experiment is realized by cascading the previous LCMS with a second one, patterned along the $y$ direction (see Fig.~\ref{fig:setup}b)). The sequence thus implements a two-dimensional (2D) periodic SU(2) transformation. The birefringence setting is ${\delta_1=\pi/2}$ and ${\delta_2=\pi}$. Figure~\ref{fig:HH-2D}a) shows the experimental far-field distribution ${\tilde{I}_{HH}(k_x,k_y)}$, from which a 2D power spectrum $P(m_x,m_y)$ is extracted. The total far-field reconstructions for a $\ket{L}$ input state are plotted in Fig.~\ref{fig:HH-2D}b), where the comparison with the experimental result is also provided. We obtain ${s_{\text{FQPT}}=88.5\%}$ and ${s_{\text{ML}}=96.7\%}$, with a total error of ${\Delta_{\text{FQPT}}=0.080}$ and ${\Delta_{\text{ML}}=0.013}$. The slight deterioration in the final prediction can be ascribed to the reduced performance of the GS algorithm in two spatial dimensions~\cite{GSmodified}. 
Figure~\ref{fig:HH-2D}c) shows the near-field amplitude ${\sqrt{I_{HH}(x,y)}}$, as reconstructed from FQPT and ML, compared with the experimental measurement. In this case, the GS algorithm runs ${N_T^{\text{2D}}=50}$ times for ${N_I^{\text{2D}}=500}$ iterations. 
In this realization, we compress the experimental images by integrating light intensity over ${11\text{ pixel}\times 11\text{ pixel}}$ regions equally distributed on the camera. This allows for both minimizing the errors due to local intensity fluctuations in the image area and keeping the computation time within the same range as the 1D experiment. The reconstructions of individual parameters are shown in Fig.~\ref{fig:HH-2D}d). An alternative visualization of the reconstructed process in terms of the local eigenvector and rotation angle is provided in Fig.~\ref{fig:setup}c). A good agreement is observed between the two predictions, with 
average fidelity $\bar{F}=91.4\%$.



\section{Discussion and conclusions}
In this work, a new technique for fast and accurate Quantum Process Tomography is demonstrated. This is accomplished via a non-interferometric scheme requiring no a-priori information on the unknown operator. In the case of complex polarization transformations, our method achieves performances very close to the standard tomography based on an overcomplete set of measurements. It offers experimental advantage, only requiring a near-minimal set of measurements, and computational speed-up, outperforming the standard approach by at least one order of magnitude. It appears naturally suitable for all experimental setups that allow for easy access to conjugate domains. For this reason, this method can also be implemented on other physical platforms, such as quantum circuits for atoms~\cite{Guidoni_1999,Gross-review} and electron beams~\cite{Harris2015StructuredQW,Grillo-electrons}.

Nevertheless, we believe that the performance of the method can be further improved by adopting optimized strategies for phase retrieval. For instance, Convolutional Neural Networks represent a promising solution~\cite{White:21,Bao2021,Ding_2022}. We also expect that FQPT can detect non-unitary evolutions~\cite{Wang:23} if equipped with some minor modifications, such as including multiple intermediate planes in the final analysis. At the same time, it would be interesting to adapt the present method to retrieve multi-photon gates and complex operations in high-dimensional Hilbert spaces. 

In principle, this procedure can also be applied to processes acting on a $m\times d$ dimensional space, having the irreducible form $\bigoplus_{i=1}^{d}\mathcal{U}^{(i)}$, with ${\mathcal{U}^{(i)}\in \text{SU}(m)}$. In this case, one must consider the decomposition of $\mathcal{U}^{(i)}$ in the generators of $\text{SU}(m)$, the generalized Gell-Mann matrices. 
Although the analytical relations between the process parameters and the measurement outcomes become significantly more complicated, the number of measurements required by FQPT will still be optimal. 
This will be investigated in successive works.


\section*{Acknowledgements}
This work was supported by the Canada Research Chair (CRC) Program, NRC-uOttawa Joint Centre for Extreme Quantum Photonics (JCEP) via the Quantum Sensors Challenge Program at the National Research Council of Canada, and Quantum Enhanced Sensing and Imaging (QuEnSI) Alliance Consortia Quantum grant.

\bibliography{bibliography.bib}

\end{document}